\documentclass[a4paper]{PoS}
\usepackage[authoryear,square]{natbib}
\bibpunct{(}{)}{;}{a}{}{,}

\newcommand\araa{ARA\&A}
\newcommand\apj{ApJ}
\newcommand\apjl{ApJ}
\newcommand\apjs{ApJS}
\newcommand\aap{A\&A}
\newcommand\mnras{MNRAS}
\newcommand\nat{Nature}

\title{Measuring Magnetic Fields Near and Far with the SKA via the Zeeman Effect}

\ShortTitle{The SKA and the Zeeman Effect}

\author{
\speaker{Timothy Robishaw}$^1$\thanks{On behalf of the SKA Cosmic Magnetism Working Group.},
James A. Green$^2$,
Gabriele Surcis$^3$, 
Wouter Vlemmings$^4$, 
A.~M.~S. Richards$^5$,
Sandra Etoka$^6$, 
Tyler Bourke$^2$, 
Vincent Fish$^7$, 
Malcolm Gray$^5$, 
Hiroshi Imai$^8$, 
Busaba Kramer$^9$, 
James McBride$^{10}$, 
Emmanuel Momjian$^{11}$,
Anuj Sarma$^{12}$, 
Albert Zijlstra$^5$
\\ 
$^1$National Research Council Canada;
$^2$SKA Organization;
$^3$Joint Institute for VLBI in Europe;
$^4$Chalmers University of Technology
$^5$Jodrell Bank Centre for Astrophysics;
$^6$Hamburger Sternwarte;
$^7$MIT Haystack Observatory;
$^8$Kagoshima University;
$^9$Max Planck Institute for Radio Astronomy;
$^{10}$University of California at Berkeley;
$^{11}$National Radio Astronomy Observatory;
$^{12}$DePaul University
\\
E-mail: \email{tim.robishaw@nrc-cnrc.gc.ca}
}

\abstract{The measurement of Zeeman splitting in spectral lines---both in
  emission and absorption---can provide direct estimates of the magnetic
  field strength and direction in atomic and molecular clouds, both in our
  own Milky Way and in external galaxies.  This method will probe the
  magnetic field in the warm and cold neutral components of the
  interstellar medium, providing a complement to the extensive SKA Faraday
  studies planning to probe the field in the ionized components.}

\FullConference{
Advancing Astrophysics with the Square Kilometre Array\\
June 8-13, 2014\\
Giardini Naxos, Sicily, Italy}

\begin{document}

\section{Introduction}

The Zeeman effect leaves its fingerprint on the circular polarization of a
spectral line.  A magnetic field threading the gas where a spectral line
for certain atomic and molecular transitions is emitted or absorbed will
cause the two senses of circular polarization to split apart in frequency.
The amplitude of the splitting between the two components is directly
proportional to the strength of the magnetic field and a constant that
depends on the atomic or molecular transition:
\begin{equation}
\Delta\nu[{\rm Hz}] = b B_{\rm los}[\mu {\rm G}] (1+z)^{-1}\,, 
\end{equation}
where $b$ is the Zeeman coefficient of the transition being observed and is
measured in Hz $\mu$G$^{-1}$, $B_{\rm los}$ is the line-of-sight magnetic
field measured in microgauss, and $z$ is the redshift of the source.
Tables of all known Zeeman coefficients can be found in \citet{heilesgmz93}
and \citet{robishaw08}.

The amplitude of the effect measured in the Stokes $V$ spectrum (defined as
IEEE right-hand minus left-hand circular polarization) towards sources in
the Galactic interstellar medium (ISM) is generally very weak compared to
the total intensity of the line measured in the Stokes $I$ spectrum because
the splitting is usually very small compared to the width of the line.
However, some Galactic maser emission lines can be so narrow that the two
circularly polarized components appear completely split; in this case, the
splitting is proportional to the total field strength rather than the
line-of-sight field \citep{crutchertghkm93}.

In this chapter we discuss the potential for using the SKA to measure the
Zeeman effect in various targets of interest in increasing order of
distance.

\subsection{Assumptions in This Chapter}

The document ``SKA1 Imaging Science Performance'' \citep{braun14} provides
parametrizations for sensitivity in spectral line and continuum modes that
are the basis for the performance estimates in this chapter. In particular
we utilize their Figure 5, which shows the spectral line sensitivity
distribution for a single deep pointing with channel width of 30 km
s$^{-1}$ and an integration time of 1000 h.  The $L$-band 1--2 GHz range
shows that for beam widths between 0.3$''$ and 300$''$, we would measure a
maximum rms of 12 $\mu$Jy, with a minimum rms of 4 $\mu$Jy at 10$''$.


\section{Molecular Clouds and Star Forming Regions in the Milky Way}

The role of magnetic fields during the formation of high-mass stars
($M>8M_\odot$) is still a matter of debate, even in the face of many recent
investigations, both theoretical and observational
\citep{seifriedbpk12,surcisvlhq13}.  Recent magnetohydrodynamical
simulations show that the collimation of outflows and the formation of
accretion discs strongly depend on the strength of magnetic fields
\citep{seifriedbpk12}.  In addition, simulations show that magnetic fields
also prevent fragmentation, reduce angular momentum, and determine the size
of H\,{\sc ii} regions
\citep{petersbkm11,hennebellecjkktt11,seifriedbpk12}. The simulations take
into account only relative values of magnetic field strengths obtained
considering ratios of magnetic energy with other types of energies (e.g.,
gravitational energy).  Therefore, no real values of magnetic field
strengths are currently used in the theoretical calculations. To provide
observational measurements of magnetic field strengths close to the massive
young stellar objects (YSOs) is of great importance.  The best probes of
magnetic fields at small scales close to YSOs are masers, in particular the
Class II methanol maser transition at 6.7 GHz and 22 GHz water masers
\citep{vlemmingsdlt06,surcisvchts11,surcisvlhq13,surcisvlgtcckk14}. Methanol
masers, which are the most numerous high-mass star formation (HMSF) maser
species, are playing a crucial role in determining the magnetic field
morphology \citep{vlemmingsstl10,surcisvlh12,surcisvlhq13}. Moreover,
\citet{vlemmingshc06,vlemmingstd11} showed that at arcsecond resolution,
methanol masers also display significant circular polarization.  In fact,
70\% of their total Effelsberg+Parkes flux-limited ($>$50 Jy) sample (51
sources) show circularly polarized emission.  This has also been confirmed
at milliarcsecond scale via interferometric observations for a smaller
number of sources \citep{surcisvlh12,surcisvlhq13}. However, the exact
proportionality between the measured Zeeman-splitting of the methanol maser
emission and the magnetic field strength is still under investigation by a
team of theoretical chemists and astrophysicists and the final results will
be published in 2015.  Current instrumentation limits the maser magnetic
field studies to up to only two dozen sources, for which only a few lines
of sight can be probed. To properly compare with models and simulations,
one would need several tens of field measurements on the individual maser
features of each source. Additionally, to properly sample different
evolutionary stages, a total of up to 100 sources needs to be observed in
detail.  Unfortunately, our ability to increase this sample is completely
limited by the currently available instruments, which allow us to measure
the very weak circular polarization signal (usually of the order
0.3\%-0.5\%) only towards methanol maser sources with flux exceeding 25 Jy
in a reasonable observation time.  Fortunately, the upcoming SKA1-MID will
allow us to investigate a much larger sample of sources.  A 30 min
integration time at a resolution of 0.1 km~s$^{-1}$---necessary to detect
the splitting---will produce an rms of 9 mJy at 6.7 GHz.  This means that
we could make 5$\sigma$ detections down to a circular polarization fraction
of at least 0.3\% towards methanol masers with flux $>$15 Jy. Therefore
with SKA1-MID we could easily measure the magnetic field strength in
hundreds of HMSF regions and consequently enlarge the investigated area of
the Milky Way.

The SKA will be invaluable in the study of Galactic masers, such as very
high-resolution follow-up of catalogues like the Methanol (and OH)
Multibeam Survey \citep{greencfabbbcetal09}. Multiple transitions provide a
chronometer with thousand-year resolution for HMSF concentrated in the
Galactic plane \citep{ellingsen07,breenecl10}. Southern hemisphere
centimeter-wave studies with resolutions of a few hundred milliarcseconds
or better are needed to complement ALMA and to map magnetic field
structures. At the other end of stellar evolution, we do not really
understand the precise mechanisms whereby cool red giants, AGB stars and
supergiants lose mass, nor how almost-spherical stars give rise to
asymmetric PNe and SNR.  The SKA and ALMA will not only map the kinematics
and magnetic fields at high resolution with masers, but, for the first
time, relate these to thermal lines and dust emission with high precision.

Evolved OH/IR stars commonly exhibit OH maser emission that probes the
outer part of their circumstellar envelopes (CSEs) where ordered magnetic
fields of a few milligauss are detectable \citep{etokad10,etokad04}.  Such
observations can trace fields even at the proto-planetary-nebula stage
\citep{etokazrml09}. Though they start their lives as spherically symmetric
objects, a large fraction of planetary nebulae are asymmetrical
objects. Although the change of symmetry might be linked primarily to
binarity, the organized magnetic fields detected at the distance where OH
masers operate in the CSEs seem to indicate the possible role of magnetism
in this morphological development.  It is anticipated that thousands of
such objects can be detected throughout the Galaxy down to a flux limit of
4 mJy with SKA1-SUR in $\sim$140~hours \citep{etoka15}.


\section{The Magnetic Field Structure of the Milky Way}

\subsection{H\,{\sc i} Absorptions Lines against Background Continuum Sources}

The magnetic field in the cold neutral medium (CNM) can be probed via the
absorption of continuum emission from compact sources at 21 cm.  This
method was successfully used by \citet{heilest04,heilest05} at the Arecibo
telescope to make an uprecedented survey of the CNM towards 41 radio-loud
sources; 20 magnetic field components were detected throughout the Milky
Way.  The limiting sensitivity of their survey was 3 mJy.  For a 10$''$
beam and 0.5 km~s$^{-1}$ channels, SKA1-MID would achieve this sensitivity
in 75 minutes.  Therefore, a census of the CNM fields throughout the
southern sky should be readily achievable.  It should be noted that FAST
will certainly make great advances in this endeavour in the northern sky
before SKA is online.  However, it should be said that single-dish H\,{\sc i}
absorption spectra are produced by subtracting an off-position spectrum and
are therefore inherently inaccurate; the possibility also arises that
polarized sidelobes can contaminate Stokes $V$ spectra for weak sources
when using this method.  These problems are mitigated by interferometric
measurements.

\subsection{H\,{\sc i} Self-Absorption}

An intriguing group of targets includes positions that show 21-cm
self-absorption \citep[][ the ``self'' prefix distinguishing this case from
  that of absorption against a background continuum source]{lig03}.  Dozens
of clouds throughout the Perseus/Taurus molecular complex show H\,{\sc i}
absorption at velocities for which OH and CO emission is seen.
\citet{mcclure-griffithsdggh06} detected H\,{\sc i} self-absorption
throughout the Riegel-Crutcher cloud, an 8$^\circ$-by-8$^\circ$ region at
the Galactic Center showing an incredible network of long, cold filaments.
When the method of \citet{chandrasekharf53} is used to investigate stellar
polarization measurements along these filaments, magnetic fields in excess
of 30 $\mu$G are inferred.  The H\,{\sc i} absorption profile has a height
of 100 K and a width of 3.5 km~s$^{-1}$.  A 30 $\mu$G field produces a
Stokes $V$ profile with peak-to-peak extent only 1.6\% of the line height,
or 1.6 K.  This can be fitted if observed with a sensitivity of 0.3 K.  The
width of the filaments is between 2$'$--5$'$, so an integration time of 2
hours per pointing would be required if observed by SKA1-MID with a 120$''$
beam.  The map would require hundreds of pointings, meaning this is best
left for SKA2 when the sensitivity will increase tenfold and the field of
view will increase twentyfold.

\subsection{Diffuse 21-cm Emission}

Diffuse 21-cm emission is seen in every direction in the sky.  This fact,
combined with the strong splitting coefficient for the 21-cm line, makes
H\,{\sc i} emission an enticing target for Zeeman studies.  This line of
work was pursued successfully using single dishes for decades before being
abandoned in the mid-1990s
\citep{heilest82,heiles88,heiles89,heiles96,heiles97,trolandh82a,
  trolandh82b,verschuur89,goodmanh94,myersggh95}. Reliable measurements
require a careful accounting of instrumental circular polarization
contributions from polarized sidelobes of the telescope.

As far as we know, nobody has successfully used an interferometer to
measure 21-cm Zeeman splitting in diffuse emission. This scenario maintains
at the VLA because its circular polarization response contains a severe
squint induced by placement of the $L$-band feed far off the symmetry axis
of each dish.  This is not impossible to model, measure, and account for,
but it is a major undertaking.  The SKA dishes will be designed such that
the secondary and feed alignment satisfy the ``Mizigutch criterion''
\citep{mizugutchay76}, which should minimize cross polarization and,
therefore, the Stokes $V$ squint.  If the circularly polarized beam can be
characterized for the SKA, then we presume that the instrumental
contribution to a Stokes $V$ measurement can be modeled and accounted
for. (Indeed, \citealt{agudo15} require circular polarization precision of
0.01\% to measure relativistic jet properties.)  The many polarization
projects that will be undertaken with the SKA guarantee that a thorough
accounting of the polarizaton properties of the array will be made, thus
enhancing the possibility that the SKA might be the first interferometer to
study the Zeeman effect in diffuse 21-cm emission.

\subsection{OH Masers Tracing the Galactic Magnetic Field}

\citet{davies74} found the amazing result that the magnetic field measured
in OH masers seemingly traces the large-scale field that the masers are
embedded in.  VLBI observations of OH masers at the highest resolutions
have demonstrated that maser Zeeman splitting shows field directions and
magnitudes which are largely coherent
\citep[e.g.][]{fishram03,vlemmingsl07}. The splitting is often replicated
in the lower resolution single-dish studies
\citep{fishraz05,szymczakg09}. Following the work of \citet{davies74},
several authors have investigated the concept of maser Zeeman splitting
tracing the Galactic magnetic field
\citep[e.g.][]{reids90,fishram03,hanz07}, finding fields consistent across
kiloparsec scales. These studies were conducted mostly with samples of
masers collated from a range of heterogeneous observations; the largest set
of systematic observations were those of \citet{fishram03}, but these were
limited to only 40 star-forming regions, all visible from the northern
hemisphere, and with only a few masers per spiral arm. Despite these
limitations, there is an implication that the magnetic fields traced by the
masers are tied to the large-scale Galactic magnetic fields, such as those
traced by rotation measures
\citep[e.g.][]{brownhgtbmdg07,vaneckbsrmgstetal11}. 

The `MAGMO' survey \citep{greengrcm14} was launched at the Australia
Telescope Compact Array (ATCA) in order to potentially map out the magnetic
field in the Galactic plane by searching for Zeeman splitting in OH
masers. MAGMO observations achieved a sensitivity of 50 mJy in 30 min of
ATCA observing with an 8$''$ beam, and consisted of targeted follow-up of
positions where 6.7 GHz methanol masers were detected. This species of
maser is an exclusive tracer of HMSF
\citep{minierenb03,pestalozzimb05,xulhpmh08,breenecgcsdv13} and as such
traces the key structural features of the Galaxy---the spiral arms, 3-kpc
arms and bar interaction. The combination of structure and magnetic field
information can be a very powerful tool for understanding the dynamics and
evolution of the Milky Way.

The logical extension of the MAGMO project would be a blind gridded survey
for OH masers in the Galactic plane using SKA1-MID. A survey covering
$-2^\circ$$<$$b$$<$$+2^\circ$ and $190^\circ$$<$$l$$<60^\circ$ would
subtend 920 sq deg, providing a sensitivity of 3.7 mJy in one month of
observing at the 18 cm OH transitions with 0.1 km~s$^{-1}$ resolution and a
10$''$ beam. The scale height for HMSF is well known and this latitude
range will capture the vast majority of star forming sources ($>$95\%).

A blind survey to lower sensitivities than achieved by the targeted MAGMO
study will bring about a larger detection rate towards regions of HMSF and
will increase the number of Zeeman detections. This will markedly improve
statistics, both within individual regions, and within Galactic structures
such as individual spiral arms. The increased number of measurements will
allow for a comparison of maser field directions with those probed via
Faraday rotation of polarized continuum sources through nearby sightlines
that contain a magnetoionic medium, which will in turn be vastly improved
in the SKA era through all-sky polarimetric surveys
\citep{johnston-hollitt15}.

An SKA-MAGMO study will explore whether the additional sensitivity that the
SKA affords will result in a larger detection rate towards regions of HMSF
or a larger number of sources per HMSF site. With the extra sensitivity we
would expect an increase in the number of Zeeman pairs/triplets, which will
provide more measurements to compare with Faraday rotation measures,
decreasing the statistical error. The additional measurements will allow
exploration of whether there is any relation to magnetic field strength for
much weaker sources, also exploring the proportion of Zeeman pairs compared
to Zeeman triplets for weaker sources.  The SKA-MAGMO study would further
enhance our understanding of the dynamics and evolution of our Galaxy
through association with precise positions of HMSF regions
\citep{green15}. 


\section{Zeeman Splitting in External Galaxies}

Masers have the highest luminosity per unit frequency of any radio source,
so it is natural to use Zeeman splitting in these beacons to measure
magnetic fields in distant galaxies.  Luckily, as we saw above in
discussing the MAGMO project, OH has a very large Zeeman coefficient and is
therefore a sensitive tracer of magnetic fields.  We discuss the
possibilities of using OH masers and megamasers as extragalactic
magnetometers.

\subsection{OH Masers in Nearby Galaxies}

SKA2 could be used to survey the local group galaxies for OH maser
emission and Zeeman splitting.  If detected, the field structure inferred
would be of great interest when comparing to MAGMO results in our own
Galaxy because our Galactic study will be confined only to the plane.

\citet{brooksw97} detected OH masers in the LMC, needing a sensitivity of
$\sim$40 mJy to obtain a 5$\sigma$ detection in Stokes $I$.  For a beam of
10$''$ and a channel resolution of 0.1 km~s$^{-1}$, the SKA1-MID would
require less than a minute to detect Zeeman splitting in such a source.  A
survey of the OH maser emission in the LMC and SMC is therefore of great
interest: any field structure detected in the OH maser distribution could
be compared directly with the fields already mapped in the SMC and LMC via
Faraday rotation \citep{maogshmsd08,gaenslerhsdmdw05}.

OH masers have yet to be detected in M31.  \citet{willett11} conducted a
search using the VLA and found nothing above a 5$\sigma$ 10 mJy limit
concluding that an order of magnitude increase in sensitivity would be
required to probe the OH masers.  This would require 10 hours of SKA1-MID
integration time with a velocity resolution of 1 km~s$^{-1}$.

M82 is the canonical starburst galaxy, located at a distance of 3.5 Mpc.
\citet{argopbmf10} used 61 hours of VLA observing with a 1$''$ beam to show
that OH masers (actually, ``kilomasers'') are seen throughout M82 with line
widths of $\sim$10 km~s$^{-1}$ and fluxes as weak as 2 mJy.  For a
5$\sigma$ detection of all the Stokes $I$ features in M82, we'd require 7
hours of SKA1-MID observing time with 1 km~s$^{-1}$ channels.  However,
unlike in the Milky Way, the OH maser lines in M82 are broad enough that
they are only partially split even for large fields of 3 mG.  The
peak-to-peak amplitude of the Zeeman feature in Stokes $V$ will be 50\% of
the line height for a splitting induced by a 3 mG field. For the weakest
detected OH maser, we can fit a Zeeman profile easily to a sensitivity
limit of 0.3 mJy.  So to detect Zeeman splitting in all observed maser
features using SKA1-MID would require 13 hours of integration.  This would
allow for a complete mapping of the field as traced by OH masers in M82.

\subsection{Megamasers}

Megamasers have been detected in hundreds of galaxies at $z \sim 0.1$, and,
with the aid of lensing, as far away as $z=2.6$
\citep{castangiaimhbrwoetal11}. Intrinsically compact both spatially
(detectable down to micro-arcsec scales) and spectrally (sub-km~s$^{-1}$),
they provide the best directly mappable tracers of high-resolution
structure in the inner few hundred parsecs of active galaxies. The early
stages of the SKA are suited to studies of the larger-scale OH masers (rest
frequency 1.67 GHz) associated with nuclear starbursts and Seyfert
galaxies. The position of individual maser spots can be measured with an
accuracy proportional to (beam size)/(signal-to-noise). For example, at a
redshift of order 0.05, MERLIN (resolution 120 mas) could resolve 10-parsec
details in Markarian 231 and 273 \citep{richardskyccwgf05,yatesrwcgfc00}
showing warped discs and orbiting mass densities of 300--900 M$_\odot$
pc$^{-3}$.

The most exciting use of OH megamasers (OHMs) has been the measurement of
\textit{in situ} magnetic fields in 15 external galaxies by
\citet{robishawqh08} and \citet{mcbrideh13}, who used the high spectral
resolution of Arecibo to separate multiple Zeeman components and compared
these with VLBI and MERLIN total-intensity lower-resolution spectra
extracted from spatially-discrete regions. The inferred magnetic field
strengths of 0.5--80 mG provide an energy density comparable to the
hydrostatic gas pressure in the masing regions, likely to be active star
formation sites.  The results from OHMs also suggest that magnetic fields
are dynamically important throughout the central starburst region, whereas
weaker magnetic fields are inferred from radio synchrotron measurements
assuming equipartition. The OHM-derived estimates are consistent
with the linearity of the far-infrared--radio correlation
\citep{mcbrideqhb14}.

The total OHM velocity span can exceed 1000 km~s$^{-1}$ but typical Zeeman
splitting detections have been associated with lines that have velocity
widths $<$20 km~s$^{-1}$ and flux densities $>$3 mJy. The Arecibo
detections required an rms flux density of 3 mJy in 0.5 km~s$^{-1}$
channels.  SKA1-MID will be able to survey all southern galaxies for OHM
Zeeman splitting; a 1$''$ beam at 1.6 GHz will reach 1 mJy sensitivity in 2
hours with 0.5 km s$^{-1}$ channels, sufficient to detect analogues of the
sources in the Arecibo sky.  If all SKA-visible Arecibo targets were
searched with this threefold improvement in sensitivity, we could expect to
double the number of galaxies with Zeeman detections in 2 days of observing
with SKA1-MID.  The southern sky has not been thoroughly searched for OHMs;
blind 21-cm surveys with the SKA are bound to catalog the southern OHM
population.  Deep follow-up observations of the southern OHM galaxies will
yield at least another doubling of the number of Zeeman-detected galaxies;
sampling the 100 brightest OHMs would require 200 hours of SKA1-MID time.
The most distant Zeeman detections occurred in OHMs at $z\sim0.2$.  The
sensitivity increase afforded by SKA2 will allow Zeeman splitting to be
probed in OHMs out past $z=1$, allowing these beacons to become invaluable
tools for studying galactic magnetic fields through cosmic time.

Previous OHM studies have relied on literature VLBI imaging, but this adds
uncertainty as the intervals between observations are comparable to the
variability timescales of compact masing clouds. SKA2 will reach the same 1
mJy rms per 120 mas beam in the same time in 25 km s$^{-1}$ channels,
allowing comparison between simultaneous high spectral- and high
spatial-resolution results.  SKA2 will extend its reach to more highly
redshifted OH masers and possibly to 22 GHz water masers, which trace
material orbiting black holes and jet-ISM interactions on sub-pc
scales. However, the splitting coefficient for H$_2$O is 1000 times weaker
than that for OH such that Zeeman splitting in water megamasers has thus
far eluded detection \citep{modjazmkg05}.


\subsection{H\,{\sc i} Absorption in Damped Ly$\alpha$ Absorbers}

Damped Ly$\alpha$ absorbing systems (DLAs) are a class of quasar absorber
in which hydrogen remains mostly neutral.  The neutral gas content of the
Universe is dominated up to redshift 5 by DLAs and the H\,{\sc i} layers
producing the absorption are considered to be the progenitors of modern
galaxies.  DLAs are perhaps the best and only sample of an interstellar
medium in the high-redshift Universe \citep{wolfegp05}.  As such, the
possibility of measuring Zeeman splitting in the 21 cm line absorption in
these systems would allow us to test the importance of the role of magnetic
fields in galaxy formation and evolution and constrain dynamo models for
the generation and amplification of magnetic fields in the early Universe.
\citet{wolfejrhp11,wolfejrhp08} describe Green Bank Telescope observations
of a DLA at $z=0.692$ towards 3C 286.  No Zeeman signature was detected in
the Stokes $V$ spectrum at 839.40 MHz down to a field limit of 17 $\mu$G.
It would take SKA1-MID 90 min to probe down to 5 $\mu$G field strengths in
this DLA with 1.5 km~s$^{-1}$ channels.  The ability to probe such systems
to lower sensitivities and therefore lower limiting magnetic fields is
tantalizing.  Because of the large redshifts of these systems, all
observations would be carried out in bands 1 and 2.  These DLAs are viable
targets for all phases of SKA deployment.


\section{Conclusions}

Measurement of the Zeeman effect is sensitivity limited.  Our ability to
use this method to probe magnetic fields in the Milky Way and beyond is
completely dependent on the development of telescopes with better
sensitivity than is currently achievable by the world's largest
observatories.  The SKA will make significant improvements in our pursuit
of measuring Zeeman splitting in spectral lines, both in emission and
absorption. These measurements will provide direct estimates of and upper
limits to the magnetic field strength and direction in atomic and molecular
clouds, both in our own Milky Way and in external galaxies.  This method
will probe the magnetic field in the warm and cold neutral components of
the interstellar medium, providing a complement to the extensive SKA
Faraday studies planning to probe the field in the ionized components.

The early phase SKA1 is estimated to yield a 50\% reduction in sensitivity,
resulting in a quadrupling of the the required integration times calculated
here. Of the possible SKA1 experiments described, only the OH megamaser
survey would be prohibitively long and require the full deployment of SKA1.


\setlength{\bibsep}{0pt}

\newcommand{\noopsort}[1]{}

\end{document}